\documentclass[]{aastex}
\usepackage{emulateapj5,onecolfloat,graphicx}
  \usepackage[hyperindex,breaklinks]{hyperref}

\def\cf{{\it cf.}}
\def\eg{{\it e.g.}}
\def\etal{{\it et al.}}

\def\ie{{\it i.e.}}

\def\SC{Paper I}

\long\def\Ignore#1{\relax}

\slugcomment{Revised version to appear in ApJ}

\shorttitle{Transient spiral modes}
\shortauthors{Sellwood \& Carlberg}

\begin{document}

\twocolumn[
\title{Transient spirals as superposed instabilities}

\author{J. A. Sellwood}
\affil{Department of Physics and Astronomy, Rutgers University, \\
    136 Frelinghuysen Road, Piscataway, NJ 08854}
\email{sellwood@physics.rutgers.edu}

\and

\author{R. G. Carlberg}
\affil{Department of Astronomy and Astrophysics, University of Toronto, \\
       Toronto, ON M5S 3H4, Canada}
\email{carlberg@astro.utoronto.ca}

\begin{abstract}
We present evidence that recurrent spiral activity, long manifested
in simulations of disk galaxies, results from the super-position of a
few transient spiral modes.  Each mode lasts between 5 and 10
rotations at its corotation radius where its amplitude is greatest.
The scattering of stars as each wave decays takes place over narrow
ranges of angular momentum, causing abrupt changes to the impedance of
the disk to subsequent traveling waves.  Partial reflections of waves
at these newly created features, allows new standing-wave
instabilities to appear that saturate and decay in their turn,
scattering particles at new locations, creating a recurring cycle.
The spiral activity causes the general level of random motion to rise,
gradually decreasing the ability of the disk to support further
activity unless the disk contains a dissipative gas component from
which stars form on near-circular orbits.  We also show that this
interpretation is consistent with the behavior reported in other
recent simulations with low mass-disks.
\end{abstract} 

]

\section{Introduction}
The origin of spiral patterns in galaxies still has no fully
satisfactory dynamical explanation \citep[see reviews by][]{BT08,
  Sell13a}.  Compelling observational evidence, both photometric
\citep{Schw76, GGF95, GPP04, ZCR09} and kinematic \citep{Viss78,
  Chem06, Shet07}, indicates that spiral patterns are
gravitationally driven density waves in the stellar disk.
\citet{KN79} and \citet{KKC11} found that the best developed, regular
patterns are observed in interacting galaxies and perhaps also those
with bars.  While the behavior in these cases may not be entirely
understood, at least the driving mechanism is clear.

The majority of disk galaxies with a significant gas component
\citep[\eg][]{Oort62} display less regular patterns, whose origin is
not so easily accounted for.  The spiral patterns may have some vague
rotational symmetry, which is predominantly 2- or 3-fold
\citep[][their Table 2]{Davi12}, but it is usually far from perfect as
arms bifurcate and/or fade with radius.  The ubiquity of the
phenomenon, taken together with the fact that simulations of isolated,
unbarred, cool collisionless stellar disks (cited below) always
manifest similar patterns, argues that there must be a mechanism for
self-excitation, which is the question we address here.

No galaxy in our hierarchical universe is truly isolated, and infalling
subhalos are predicted to bombard the outer halo of every galaxy
\citep[\eg][]{Boyl10}.  Since tides can excite spiral responses, it is
possible some patterns are excited by halo substructure
\citep[\eg][]{Dubi08}.  But the inner halos of galaxies, where fragile
thin disks reside, are quite smooth \citep{Gao11} and even large
subhalos, such as that which hosted the Sagittarius dwarf galaxy
\citep{Belo06}, can be severely tidally disrupted before perturbing
the disk \citep{Purc11}.  We argue here that spiral patterns appear so
readily as self-excited instabilities that disk responses to diffuse
sub-halo perturbations probably give rise to a minority of spirals.

Continuously changing recurrent transient patterns have been reported
over many years from simulations of isolated, unbarred disk galaxy
models \citep[\eg][]{HB74, SC84, Rosk08} and no qualitatively
different behavior has emerged as the numerical quality has improved.
Claims of long-lived spiral modes have not proven to be reproducible
\citep{Sell11}.  Spiral activity fades over time as stellar random
motions rise due to scattering by the fluctuating spiral patterns, but
a reasonable amount of gas infall and dissipation can prolong spiral
activity apparently indefinitely \citep[\eg][]{SC84, CF85, Toom90},
which is also consistent with modern galaxy formation simulations
\citep[\eg][]{Ager11}.

Here, we finally address the issue that was left unexplained in
\citet[][hereafter \SC]{SC84}, namely the nature of the spirals that
arise in such simulations.  In a follow-up study to our original work
\citep{Sell89}, we reported that the continuously changing patterns
appeared to result from the superposition of a few longer-lived waves,
each of which had well-defined frequencies and shapes and lasted for
between five and ten full rotations of the pattern.  These properties
are consistent with them being modes, or standing waves, although they
did not last indefinitely.  Here we provide stronger evidence and
propose a mechanism for this interpretation, using simulations of
altogether higher quality than those in our original study.

%\newpage
\section{Simulations}
Computational resources available at the time required that the
simulations presented in \SC\ employed typically just $2\times10^4$
particles whose motion was confined to a plane.  Here we continue to
employ the same basic physical model, with a fraction of the total
mass in a disk represented by particles, while the remaining central
attraction comes from a rigid mass distribution, but we present
results with greatly increased numbers of particles and also full
three-dimensional (3D) motion.

We mimicked dissipation in some of our models in \SC, but we do not
attempt to do so here in order to focus on purely collisionless
dynamics.  All the models reported below therefore heat as a result of
spiral activity \citep{CS85} until the disk becomes featureless after
some number of rotations.

\subsection{Mass Model}
The radial surface density profile of the disk has the form we
employed in \SC
\begin{equation}
\Sigma(R) = 1.563 {M \over a^2} \left(1+{20 R \over a}\right)^{-{3/4}}
\left(1+{R \over 5 a}\right)^{-2},
\label{eq.surfd}
\end{equation}
and a circular velocity curve
\begin{equation}
V(R) = 1.956 V_0 {R \over a} \left(1+{2R \over
  a}\right)^{-{7/8}} \left(1+{R \over 5a}\right)^{-{5/8}},
\label{eq.rotcur}
\end{equation}
where $a$ is a length scale, and $M$ a mass unit; thus the velocity
unit and dynamical time are, respectively, $V_0 = (GM/a)^{1/2}$ and
$\tau_0 = (a^3/GM)^{1/2}$.\footnote{The bizarre proportionality
  constants in eqs.~(\ref{eq.surfd}) and (\ref{eq.rotcur}) are needed
  to reconcile the units used in \SC\ with those used here.}
Expressions (\ref{eq.surfd}) and (\ref{eq.rotcur}) are not perfectly
self-consistent, but a numerical solution with softened gravity for
the central attraction of the surface density (\ref{eq.surfd}) gives a
rotation curve that is quite well fitted by the function $V(R)$ in
eq.~(\ref{eq.rotcur}).  In \SC, we devised this unconventional model
to meet two objectives: (1) to produce a rotation curve that
superficially resembled that of a large Sc galaxy, and (2) to yield a
model that developed spirals all over the disk, since we had found
that quasi-uniform rotation in the inner disk prevented spirals from
developing there.

We employ this model again in the present paper for comparison to
our earlier work.  Although the disk mass profile is not the usual
exponential, it is not unreasonable; our conclusions seem quite
general, and should hold in any plausible model of a differentially
rotating disk.  We continue to employ a rigid halo principally because
our focus is on spiral dynamics.  While it remains to be demonstrated
explicitly, we believe that a responsive halo would have very little
affect on the spiral evolution; bars can lose angular momentum to
halos through resonant interactions \citep{TW84}, but spirals have
both lower amplitude and shorter lifetimes than do bars and, hence,
couple much more weakly to a halo.  We showed in \SC\ that gas was
important to prolong the lifetimes of the patterns, but omit this
aspect also.  In summary, the simplified experiments presented here
have allowed us to develop an understanding of the driving mechanism
for spirals in the stellar disk only, which seems to be the key part
according to the observational evidence reviewed in the introduction.

We use a cubic function to taper the surface density smoothly to zero
over the range $6.5a < R < 7.5a$, leaving a total disk mass of $\simeq
3.4M$ with the half-mass radius being $R_e \simeq 2.75a$.  The
circular speed has a broad maximum of $V \simeq 0.8V_0$ around $R\sim
3a$, and a rotation period at $R=R_e$ is $\sim 20$ dynamical times.
Since the scaling to physical units is arbitrary, we here set
$G=M=a=1$; one possible scaling is to choose $a = 3\;$kpc and $\tau_0
= 10\;$Myr, leading to $V_0 \simeq 293\;$km~s$^{-1}$ and $M \simeq 6.0
\times 10^{10}\;$M$_\odot$.

In practice, we use a disk with a fractional surface density $f$, of
that given in eq.~(\ref{eq.surfd}), while the contribution from the
rigid matter gives $V^2(1-f)/r$ to the central attraction, with $V$
given by eq.~(\ref{eq.rotcur}).  In \SC, our simulations started with
$f=0.3$ to inhibit vigorous bar-forming instabilities in the disk.

We start all simulations with $Q=1$ at all radii, where
\begin{equation}
Q(R) = {\sigma_R \over \sigma_{R, \rm crit}}, \quad{\rm with}\quad
\sigma_{R, \rm crit} = {3.36 Gf\Sigma(R) \over \kappa(R)}.
\label{eq.Qdef}
\end{equation}
\citep{Toom64}.  Here $\sigma_R(R)$ is the local radial velocity
dispersion of particles in the disk and $\kappa$ is the usual Lindblad
epicycle frequency.  We set the azimuthal dispersion and mean orbital
speed using the Jeans equations in the epicycle approximation.  In 3D
models, we distribute particles in $z$ using a Gaussian of width
$z_0=0.05$, and construct vertical equilibrium by solving the 1D Jeans
equation in the vertical restoring force computed from the particles.
Since the disk mass is low, this procedure yields models that are
close enough to equilibrium that no initial relaxation is required.

We generally compute the evolution for $500\tau_0$.  We use a basic
time step $\delta t = 0.05\tau_0$, and the time step is increased by
successive factors of two in three annular zones of greater radii.  In
2D, we adopt Plummer softening with softening length $\epsilon=0.05$
and a grid with $128$ radial spokes and $96$ logarithmically spaced
rings.  In 3D models, we employ a cubic spline law \citep{ML85} with
$\epsilon=0.04$ and grid of $96 \times 128 \times 125$ mesh points.

Our numerical techniques are described in full detail in {\tt
  http://www.physics.rutgers.edu/$\sim$sellwood/code/\hfil\break manual.pdf}.

\subsection{Two-dimensional Models}
\label{sec.2D}
We first present a series of simulations with $f=0.3$ and other
physical properties exactly as we used in our first model from \SC --
but which have increasing numbers of particles.  All the models
manifested spiral activity resembling that in our earlier paper.

Initial spiral activity is seeded by shot noise, which has an
amplitude that scales as $N^{-1/2}$.  While chance leading wave
components in the spectrum of shot noise give rise to responses that
are swing-amplified by a substantial factor \citep{GLB65, Toom81}, we
showed in \SC\ that the expected noise level was below the measured
leading signal even with $N=2 \times 10^4$, and is correspondingly
lower still in the present experiments with much greater numbers of
particles.  Spiral activity at the amplitude we observe must be
boosted either by feedback creating unstable modes or by non-linear
effects or both, as we discuss in \S\ref{sec.modes}.

\begin{figure}[t]
\begin{center}
\includegraphics[width=.8\hsize]{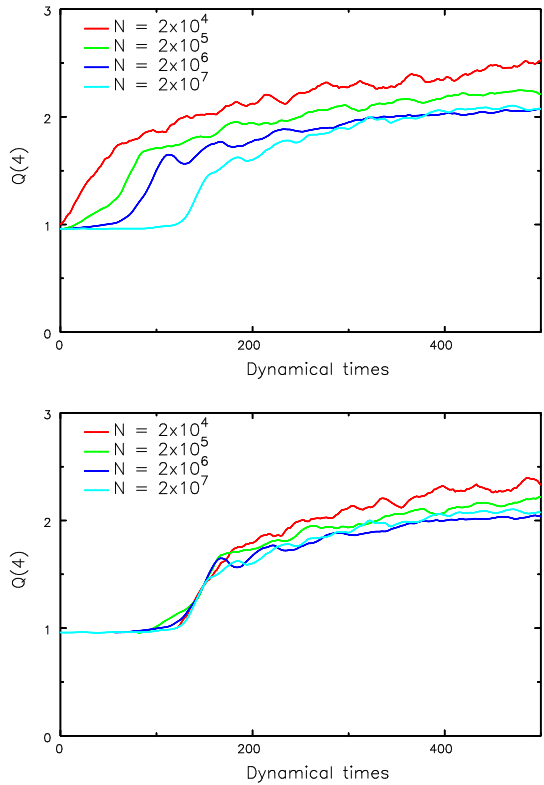}
\end{center}
\caption{Time evolution of $Q$ at $R=4$ in four simulations
  with increasing numbers of particles, still confined to move in a
  plane.  In the lower panel, the curves have been shifted
  horizontally so that all pass through $Q=1.4$ at the same moment,
  which required larger time shifts in the experiments with smaller
  $N$.  Note that a rotation period at the half-mass radius is 20
  dynamical times.}
\label{fig.heat2D}
\end{figure}

Fig.~\ref{fig.heat2D} shows the time evolution of $Q$ at a radius of
$R=4$ measured from the simulations.  Spiral activity heated all the
models, but as we employed more particles, the spirals took
increasingly long to develop.  To compensate for this delay, which we
account for in \S\ref{sec.modes}, we slide the curves from the smaller
$N$ models to the right in the lower panel so that they all coincide
as $Q$ rises through $1.4$, which reveals that the $N=2 \times 10^4$
model, as used in \SC, heated a little more rapidly and to a slightly
higher value of $Q$ than did the larger $N$ models.

However, the heating rates are all very similar once spiral activity
gets going, indicating that rapid heating is not caused by two-body
relaxation, as some authors \citep[\eg][]{Fuji11} have suggested.  We
offer four additional reasons to reject their conjecture.  First, we
noted in \SC\ that heating was much less rapid in the inner disk of a
Kuzmin-Toomre model, where shear is weak and spiral activity was
insignificant, indicating that heating was caused by the spirals and
not by relaxation.  Second, the same code has been used with similar
particle numbers to reproduce the normal modes expected from stability
analyses of smooth stellar fluids \citep{SA86} which would be
impossible if relaxation were rapid.  Third, the code reproduced the
behavior in a solution of the collisionless Boltzmann equation
\citep{Inag84}.  Finally, relaxation in 2D systems differs
fundamentally from that in 3D \citep{Rybi72, Sell13b}, and is readily
controlled by force softening.  The slower heating in the experiments
by \citet{Fuji11} and others is accounted for in \S\ref{sec.slowheat}.

See \S\ref{sec.modes} for an explanation of the origin of spirals and
\S\ref{sec.discuss} for further discussion of related work by other
authors.

\begin{figure}[t]
\begin{center}
\includegraphics[width=\hsize]{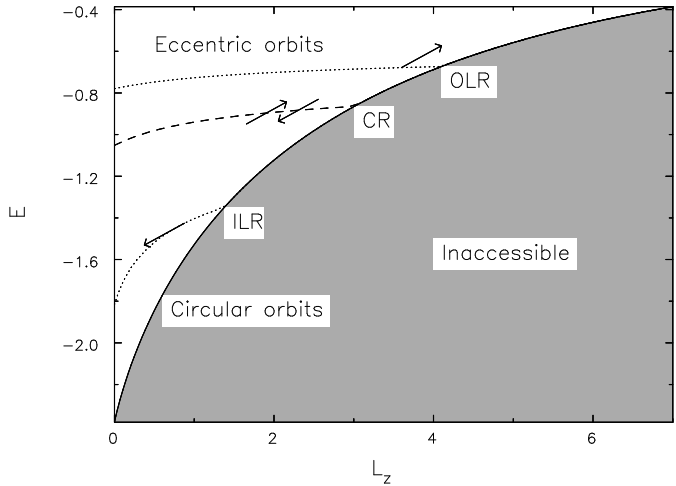}
\end{center}
\caption{Lindblad diagram showing specific energy, $E$, as a
  function of angular momentum, $L_z$, for particles in our Sc model.
  Eccentric orbits lie in the region above the solid curve, which is
  for circular orbits.  The dashed and dotted curves show the loci of
  respectively corotation and the two Lindblad resonances for an $m=3$
  disturbance with $\Omega_p=0.07$.  The arrows mark possible
  scattering vectors at the three principal resonances.}
\label{fig.lindblad}
\end{figure}

\subsection{Spiral Heating}
In all cases, the model heated to $Q \ga 2$ in a few disk rotations,
and the increased random motion caused spiral activity to fade, as
reported in \SC, and by others, in models without cooling.  Because
the dynamical time scale is shorter near the center, the inner disk
heated more rapidly than the outer.

\begin{figure*}[t]
\begin{center}
\includegraphics[width=.8\hsize,angle=270]{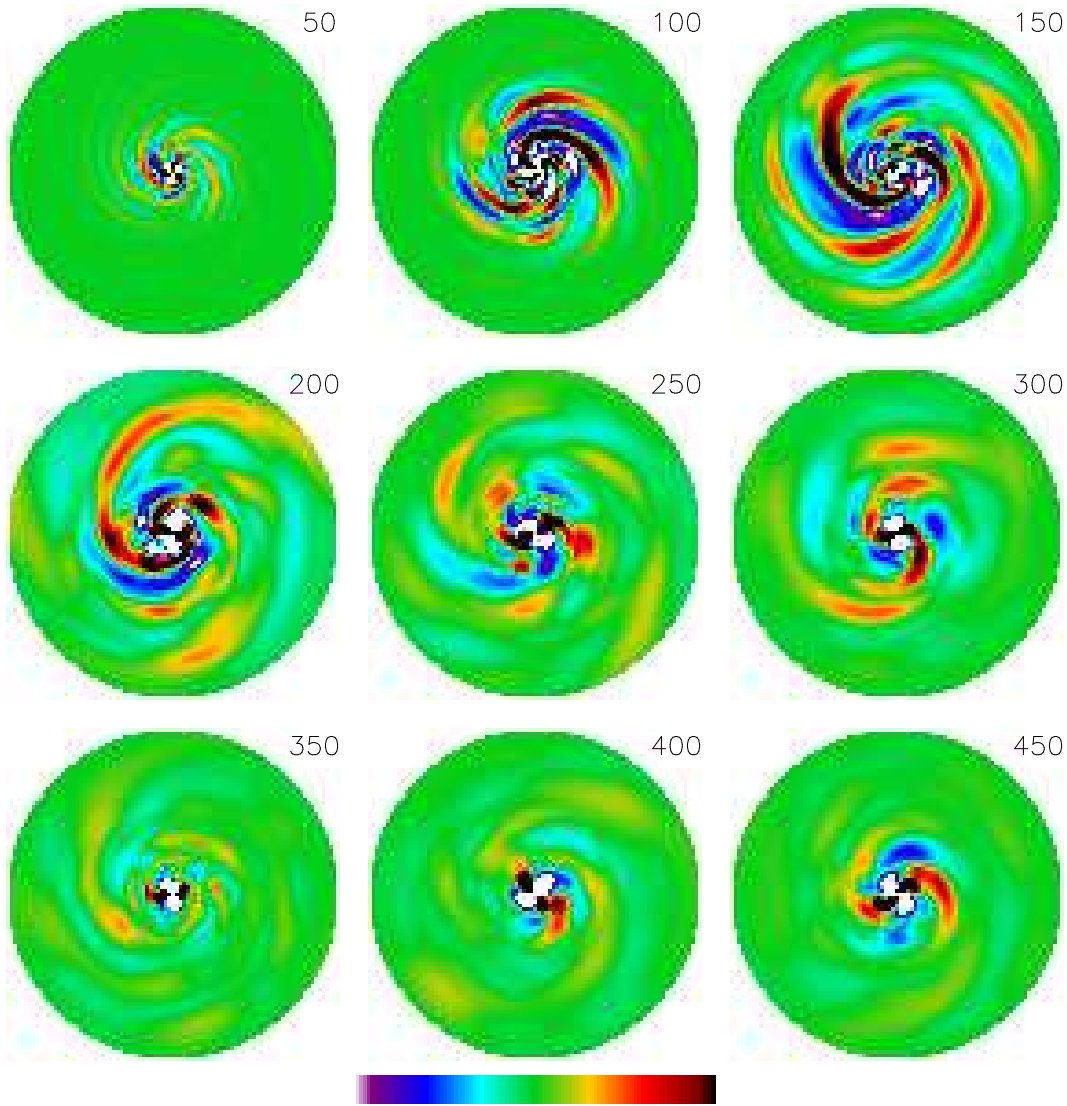}
\end{center}
\caption{Surface density of particles in the $N=2 \times 10^8$
  simulation at the times indicated.  The color scale, which ranges
  over $\pm0.5$, indicates the density relative to the mean at each
  radius; values outside this range are black or white.  The radius of
  each circle is 8 disk scales.  Notice that the spirals develop quite
  rapidly in the center, then spread outward before fading more
  slowly as the disk heats, while a small bar develops in the center.}
\label{fig.contrast}
\end{figure*}
%\clearpage \break \noindent  

Scattering at the Lindblad resonances causes an irreversible increase
in the level of random motion of the scattered particles \citep{CS85}.
Since Jacobi's integral, $I_J = E - \Omega_pL_z$, is a conserved
quantity in a non-axisymmetric potential that rotates steadily at rate
$\Omega_p$ \citep{BT08}, changes of angular momentum $\Delta L_z$ and
specific energy are related as $\Delta E = \Omega_p\Delta L_z$.  Thus
stars are scattered along trajectories of fixed slope $\Omega_p$ in
the Lindblad diagram, shown in Fig.~\ref{fig.lindblad}, and lasting
changes occur only at resonances \citep{LBK72}.  Note that the
scattering vectors at the Lindblad resonances (dotted curves) are
angled away from the slope of the circular orbit curve and scattered
particles therefore gain energy of non-circular motion.  Random
motions rise everywhere because the simulations manifest multiple
patterns with resonances distributed widely over the disk
(\S\ref{sec.waves}).

In \SC\ we demonstrated that a modest rate of addition of new
particles on near-circular orbits, in a crude attempt to mimic
accretion of cold gas that was immediately converted to star
particles, was sufficient in this model to maintain spiral activity
``indefinitely.''  Spiral activity has also been shown to persist in
simulations that employ other methods of cooling \citep[\eg][]{CF85,
  Toom90, Rosk08, Ager11}.  The persistence of spiral activity in
cooled simulations provides an attractive explanation for the strong
correspondence between the presence of spirals and gas in real
galaxies \citep{Oort62}.

\subsection{3D models}
\label{sec.3D}
We next present simulations of the same mass models that allow full 3D
motion for the disk particles.  In addition, these models have
$f=0.4$, making the disk slightly more massive.

Employing $N=2 \times 10^4$ particles, while almost enough in 2D, is
woefully inadequate once 3D motion is allowed, since the disk thickens
excessively due to 2-body relaxation \citep{Sell13b}.  We therefore
present results from experiments with $2 \times 10^5 \leq N \leq 2
\times 10^8$.

\begin{figure}[t]
\begin{center}
\includegraphics[width=.8\hsize]{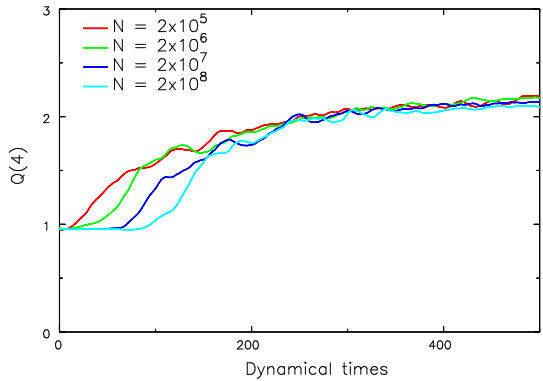}
\end{center}
\caption{Time evolution of Toomre's $Q$ at $R=4$ in simulations
  in which 3D motion is allowed and $f = 0.4$.  The four curves show
  results with different numbers of particles.}
\label{fig.heat3D}
\end{figure}

Fig.~\ref{fig.contrast} shows the perturbed surface density at a few
times in the simulation with $N=2 \times 10^8$, showing that the
evolution resembles that of the uncooled model in Paper I, even though
the particle number was increased by four orders of magnitude and 3D
motion is allowed.  Fig.~\ref{fig.heat3D} shows the time evolution of
$Q(4)$ in these 3D models for which there is again a mild
$N$-dependence.  Once again, the onset of heating is more delayed as
$N$ rises because spiral activity takes longer to get going.

\Ignore{However, the later $Q$ values are not ordered by $N$, in
particular, $Q(4)$ surged to a high value when $N = 2 \times 10^6$
because spiral activity, which is stochastic, was unusually vigorous
in the early evolution of this model.  The evolution of $Q$ in a
identical model with a different random seed did not show this early
surge.}

\subsection{Power Spectra}
\label{sec.waves}
We use Fourier transforms to compute the gravitational field from the
distribution of particles assigned to our polar grid at every step.  We
save the vertically integrated azimuthal Fourier coefficients of this
mass distribution at each grid radius at regular intervals, typically
40 time steps, or $\Delta t = 2$.  Fourier transformation in time of
these coefficients yields the power as a function of frequency, for
each sectoral harmonic, $m$, and radius $R$.  The duration of the
simulations is long enough, and the data saved at sufficiently frequent
intervals, that we can subdivide the time range in order to compare
the spectra from different periods of evolution.

\begin{figure*}[t]
\begin{center}
\includegraphics[width=.6\hsize,angle=270]{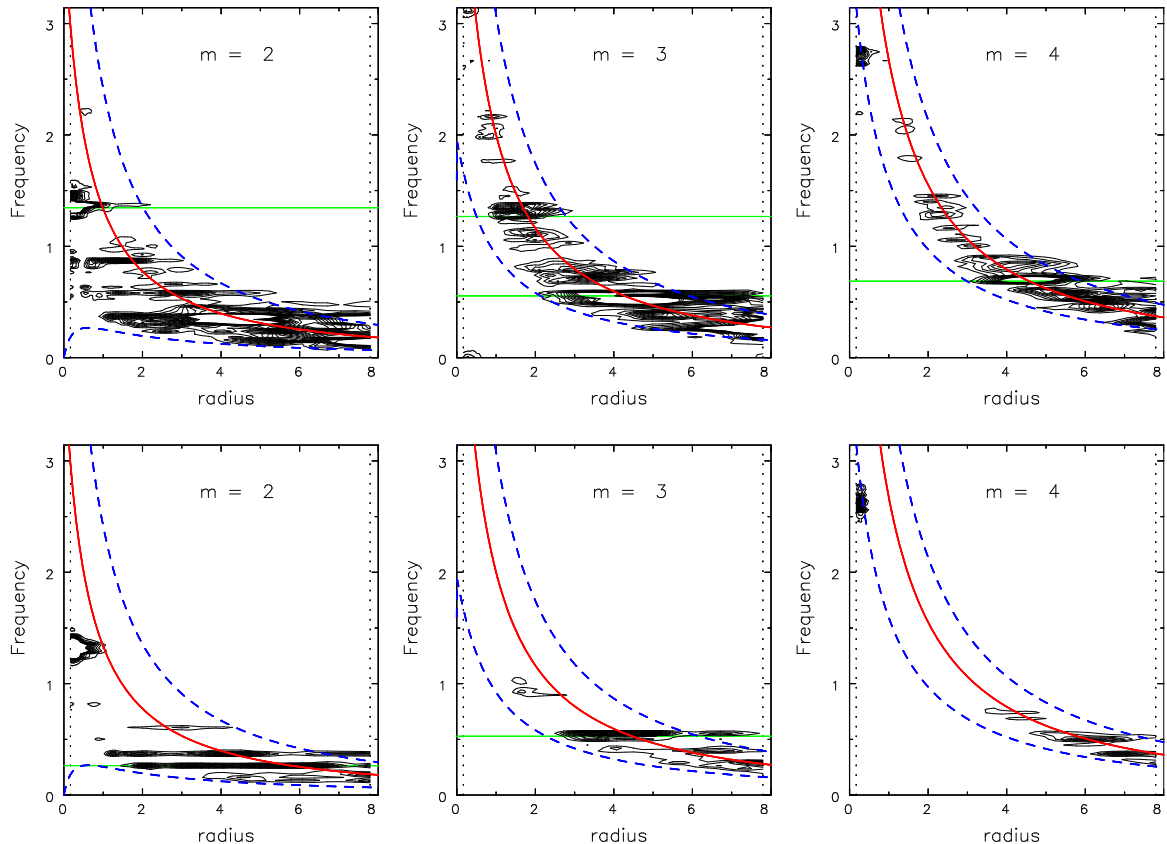}
\end{center}
\caption{Contours of power as functions of radius and frequency for
  different sectoral harmonics $m$ in our 3D model.  The top row is
  for the first half of the evolution, $22 \leq t \leq 260$, the
  bottom row shows the second half $262 \leq t \leq 500$.  In each
  panel, the solid (red) curve marks $m\Omega_c$ and the dashed (blue)
  curves $m\Omega_c \pm \kappa$.  Each horizontal stripe indicates a
  coherent wave with a well-defined angular frequency, $m\Omega_p$,
  over the time interval.  The green horizontal lines mark the
  frequencies of the waves illustrated in Fig.~\ref{fig.spirals}.}
\label{fig.pspectra}
\end{figure*}

The top (bottom) rows of Fig.~\ref{fig.pspectra} present power spectra
for three sectoral harmonics from the first (second) half of the
evolution of our 3D model with $N=2 \times 10^6$.  The contours are of
power as functions of radius and frequency, the solid curve shows
$m\Omega_c$ and the dashed curves $m\Omega_c \pm \kappa$, where
$\Omega_c(R)$ is the angular frequency of circular motion and
$\kappa(R)$ is the Lindblad epicyclic frequency.  Each horizontal
ridge indicates a coherent density wave of frequency $m\Omega_p$,
where $\Omega_p$ is the pattern speed, that persisted for some time
and extended over a range of radii.  Plots from our larger $N$
simulations are remarkably similar, although the precise frequencies
of the ridges differ.

The ridges typically have the largest amplitudes near corotation and
generally extend roughly as far as the Lindblad resonances, as
expected from the locally derived dispersion relation for spiral waves
\citep{BT08}.  In some cases, the wave outside corotation (CR) is very
weak.  Note that the high-frequency features near the center in the
right panels ($m=4$) have twice the frequencies of the stronger inner
features at $m=2$ (left panels), and are therefore not independent.

In the early part of the evolution, these coherent waves are strongest
in the inner disk, but activity spreads outward over time, while it
weakens in the inner disk, which is heated more rapidly as a result of
the spiral activity.  Movies of the time evolution of the power spectra
from all our models reveal that each density ridge typically lasts for
between 5 and 10 rotations at its corotation radius.

\begin{figure*}[t]
\begin{center}
\includegraphics[width=.6\hsize,angle=270]{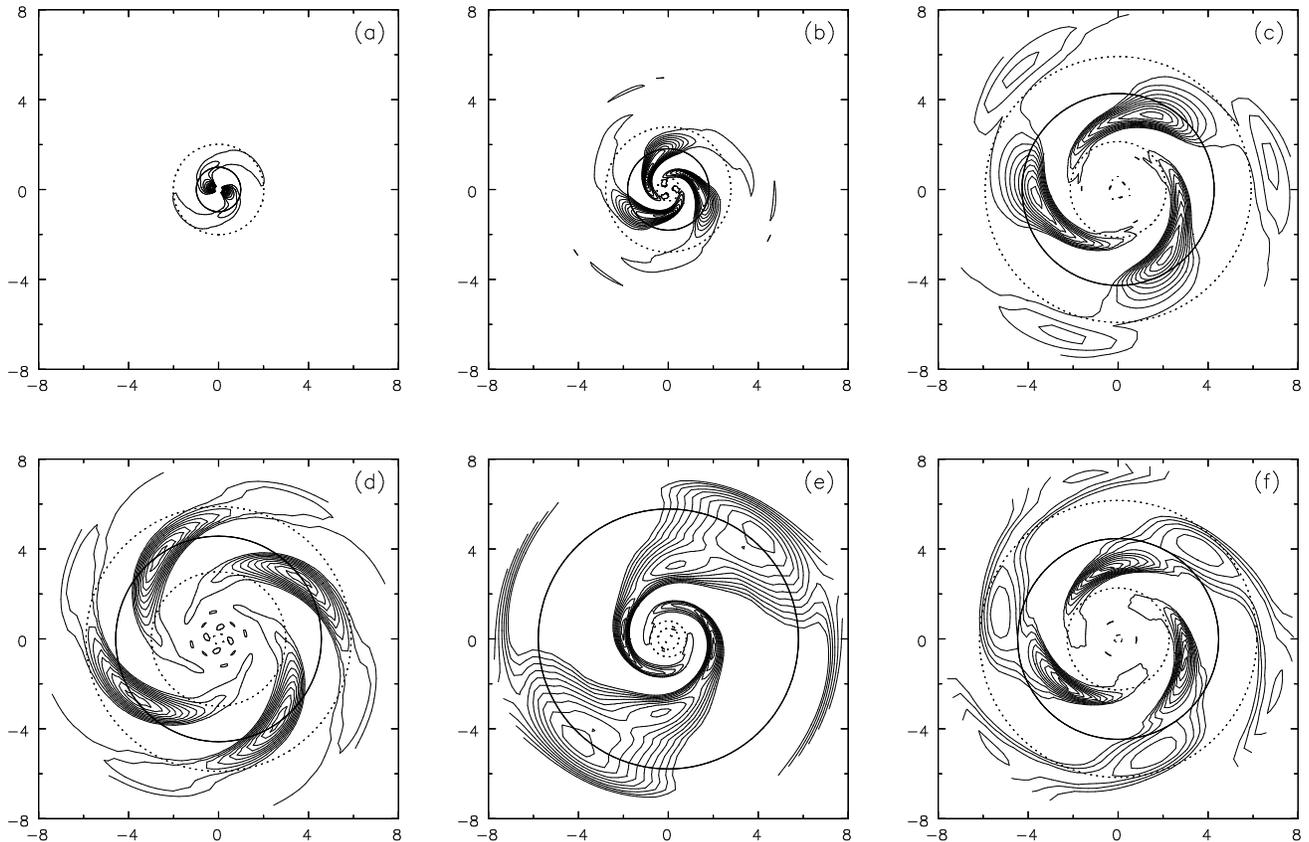}
\end{center}
\caption{Projected shapes of the six long-lived waves marked with
  green lines in Fig.~\ref{fig.pspectra}.  Contours are of positive
  overdensity only and the solid circle marks corotation, while the
  dotted circles mark the Lindblad resonances.  Panels (a) through (d)
  are from the first half of the evolution while (e) and (f) are from
  the later part.}
\label{fig.spirals}
\end{figure*}

For each frequency in Fig.~\ref{fig.pspectra}, the temporal Fourier
transform yields the amplitude and phase of the wave at the initial
moment.  From this information, we can draw the shapes of the waves
that give rise to the ridges in the power spectra.
Fig.~\ref{fig.spirals} shows the projected shapes of six waves that
were marked with green horizontal lines in Fig.~\ref{fig.pspectra},
with the corotation and Lindblad resonance radii for each drawn
by solid and dotted circles, respectively.  The wave in panel (a) is
probably a bar mode, while each of the others has a trailing spiral
form with peak amplitude near corotation; most have a more tightly
wrapped feature near the inner Lindblad resonance (ILR).

\vfill\eject
\subsection{Quiet Starts}
The above results were obtained from simulations in which the initial
azimuths of the particles were selected at random, causing the initial
behavior to be driven by the shot noise from the random positions of
the particles.  We have also run several separate 2D quiet start
simulations in order to identify instabilities of the {\em smooth}
initial disk.

A quiet start (temporarily) suppresses activity excited by particle
noise, since particles are initially spaced evenly around rings and
disturbance forces are restricted to a single sectoral harmonic $m$.
Provided the number of particles on each ring is $>2m+1$, forces from
the ring are those of a mildly distorted uniform ring, allowing
disturbances to grow by many $e$-folds before they saturate.  Fitting
the evolution of the amplitude and phase of the density variation in
the linear regime yields the pattern speed and growth rate of dominant
unstable mode(s) at each $m$; see \citet{SA86} for a full description
of the technique.

With $m=2$ only, we found a very slowly growing mode with $\omega
\simeq 0.71 + 0.03i$; the higher pattern speed of the bar mode in the
noisy start model (Figures 5 and 6a) is probably seeded by shot noise.
As the growing mode had an open bisymmetric spiral form with no ILR,
it qualifies as a bar-forming cavity mode of the kind described in
\S\ref{sec.modes}.1; the growth rate is low because the disk is
sub-maximal.  Because it grew so slowly, it did not produce a visible
bar until $t \simeq 480$ in the quiet start run.  We found a much
fiercer instability when forces were restricted to $m=3$ only: the
measured frequency was $\omega \approx 2.1 + 0.13i$.  This appeared to
be another cavity-type mode, since its rotation rate is high enough to
avoid an ILR.  But even this fierce mode is outgrown, in noisy start
simulations, by other features at larger radii as shown in the top
middle panel of Fig.~\ref{fig.pspectra}.  There were no credible
linear modes with $m=4$ only.

Note that we did not find any evidence for outer edge modes
\citep{Toom81, PL89} in any of these quiet start simulations, although
it is possible that one could be present at each $m$.  Our tapered
edge would cause any instability to grow too slowly to be detected
and, if present, a mild instability would not affect the inner disk
because the orbit time-scale is so much longer near the disk edge
($R_{\rm max}=7.5$).

\section{Nature of the waves}
\label{sec.modes}
With the principal exception of the mode in Fig.~\ref{fig.spirals}(a)
that gives rise to a small, persistent bar, we contend that almost all
the ridges in the power spectra resulted from {\it transient spiral
  modes} of the form shown in the other panels of
Fig.~\ref{fig.spirals}.  Here we explain this interpretation more
fully, and provide supporting evidence.

\subsection{Swing-amplified Modes}
The famous transient spiral presented by \citet[][his Fig.~8]{Toom81},
reproduced as Fig.~6.19 in \citet{BT08}, illustrates
swing-amplification of a {\it wave packet}.  The evolution is not that
of a mode because the wave neither maintains a fixed shape nor grows
indefinitely.  Instead it is a {\it response\/} to a particularly
provoking initial disturbance.

After this transient swings from leading to trailing, accompanied by
strong amplification, the spiral wave travels radially away from
corotation at the group velocity \citep{Toom69},\footnote{Strictly
  this is the direction of wave propagation on the short-wave branch
  of the dispersion relation only \citep{LS66, BT08, Sell13a}; the
  group velocity is in the opposite direction on the long-wave branch.
  However, the long-wave branch probably does not exist except in
  disks of very low mass \citep[see the Appendix of][]{LBK72}.}
advecting wave action, or angular momentum, across the disk.  In a
stellar disk with smoothly varying surface density and random motion,
as was adopted for this calculation by Toomre, the wave propagates
without significant losses as far as the Lindblad resonances, which
are the only locations where the wave action can be absorbed by
collisionless particles \citep{LBK72, Mark74}.

Modes, on the other hand, are standing waves between two reflecting
ends, as is familiar from guitar-strings and organ pipes.  If the
in-going trailing wave can reflect off some feature, or the center, it
becomes a leading wave with a group velocity directed back toward
corotation.  The outgoing leading wave incident on corotation is
``super-reflected,'' or swing-amplified, allowing a closed cycle with
amplification.  A growing mode is a continuous wavetrain that
propagates around this cycle.  As usual, only the frequencies for
which the phase closes can give rise to standing waves.

Modes of this type are termed cavity modes.  Reflection off the center
is the mechanism for the bar-forming instability proposed by
\citet{Toom81}.  Here we suggest that reflections can also occur where
previous disturbances have created features by resonant scattering in
an initially smooth disk.\footnote{\citet{Mark77} also proposed a
  feedback loop through reflection of waves off a feature in the inner
  part of a low-mass disk, but the $Q$-barrier he invoked caused more
  of a refraction of trailing waves from the short- to the long-wave
  branch of the dispersion relation.}

\subsection{Impedance changes}
Wave particle interactions at Lindblad resonances scatter particles to
more eccentric orbits and different angular momenta
(Fig.~\ref{fig.lindblad}).  It is significant, as noted by
\citet{Sell12} for $m=2$ waves in the Mestel disk, that scattering at
the ILR is in a direction in the $(L_z,E)$ plane that is almost along
the locus of the ILR, and therefore each scattered particle stays on
resonance causing large changes to its orbit from even quite mild
spiral patterns.  Furthermore, the ILR is localized so that only a
small fraction of the particles are affected by it.  This important
feature of ILR scattering is quite general (Fig.~\ref{fig.lindblad} is
drawn for an $m=3$ wave in our Sc model), and contrasts with the
behavior at the OLR where small changes are shared by many particles
that are each quickly scattered off resonance.

The particles scattered at an ILR of one disturbance have increased
radial excursions over a highly localized range of angular momenta.
(We show this to be the case in our models in Fig.~\ref{fig.actall}
below.)  The effect of this locally increased random motion is that
part of the disk will respond less cooperatively to subsequent waves
as they travel across the disk, with important consequences.  It is
the abrupt change to the density of particles in action space that
matters for spiral dynamics, which is not blurred by the increased
epicyclic excursions.

\citet{Toom69} applied the group velocity of a wave packet
\begin{equation}
v_g = {\partial\omega \over \partial k},
\end{equation}
to spiral waves, for which $\omega \equiv m\Omega_p$ and $k$ are the
pattern speed and radial wavenumber of the wave, respectively.  The
radial speed at which angular momentum is advected, $v_g$ can be
computed from the dispersion relation for spiral waves in a stellar
disk, which is (in the WKB approximation)
\begin{equation}
(\omega - m\Omega)^2 = \kappa^2 - 2\pi G\Sigma {\cal F}(s,\chi) |k|,
\end{equation}
\citep{LS66, BT08, Sell13a}.  Here, $\Omega$ and $\kappa$ are the
usual circular and epicycle frequencies in the axisymmetric potential,
while ${\cal F}(s,\chi)$ is the factor by which the responsiveness of
the disk to the spiral potential is reduced compared with that were
the disk to have no random motion.  It depends on two properties of
the stellar distribution: the ratio of the forcing frequency
experienced by a star to its natural frequency, $s \equiv |\omega -
m\Omega|/\kappa$, and $\chi \equiv \langle v_R^2 \rangle k^2 /
\kappa^2$, which is the square of the ratio of the typical sizes of
the stellar epicycles ($\propto \langle v_R^2 \rangle^{1/2}/\kappa$)
to the wavelength of the wave ($\propto |k|^{-1}$).  When $\chi$ is
large, ${\cal F} \ll 1$ because the unforced epicyclic amplitude of
most stars is larger than the radial wavelength of the disturbance,
and the weak supporting response arises mainly from the small fraction
of stars near the center of the velocity distribution.

In our situation, there is an abrupt change to the sizes of the
epicycles of stars that causes a corresponding abrupt change to $\cal
F$.  To understand how this affects the transmission of a spiral wave,
it is useful to introduce the concept of impedance, familiar from
electrical circuits or, more usefully, acoustic waves \citep[see
  \eg][]{Fren71}.  The impedance in a stretched string, for example,
is defined as $Z = F_y/v_y = (T\mu)^{1/2}$, where $F_y$ and $v_y$ are,
respectively, the restoring force and displacement velocity in the
vibration normal the $x$-direction of the equilibrium string, $T$ is
the tension and $\mu$ the mass per unit length.  When two strings with
differing $\mu$ are joined, the difference of impedance determines the
reflection and transmission of waves across the join.  For example, a
wave in a heavy string is perfectly reflected at a join to a massless
string.

Returning to spiral waves, the displacement velocity for spiral waves
in the local approximation is
\begin{equation}
\bar v_{Ra} = - {\omega - m\Omega \over \kappa^2 - (\omega - m\Omega)^2}
k\Phi_a {\cal F}
\end{equation}
\citep[][eq.~6.58]{BT08}, the impedance, therefore, is $Z = k\Phi_a /
\bar v_{Ra} \propto {\cal F}^{-1}$, with the frequency factor
dependent only on the properties of the axisymmetric potential and the
pattern speed of the spiral.  Thus, when a spiral wave traveling
across a disk encounters an abrupt change to $\cal F$, it will be
partly reflected and partly transmitted.  In particular, a large
decrease in $\cal F$, previously created by resonance scattering from
an earlier wave, will be highly reflective.

\subsection{Emergence from Noise}
\citet{Sell12} studied the emergence of true instabilities from noise
in highly restricted simulations of the half-mass Mestel disk with
differing numbers of particles; motion was confined to a plane and
perturbing forces arose from a single sectoral harmonic.  He reported
that changes caused by ILR scattering by earlier disturbances led to
enhanced activity in the disk, presumably because the impedance change
caused partial reflection of later in-going trailing waves.  He found
that the level of spiral activity exponentiated rather slowly until
the reflections became strong enough to trigger an unstable global
spiral mode with a much higher growth rate.

\Ignore{Sellwood \& Evans (in preparation) present further evidence
for the inner reflection of trailing to leading waves from a feature
in a similar model that was put it by hand.}

Our present simulations are slightly more general than those of
\cite{Sell12}, because they allow 3D motion and include multiple
sectoral harmonics in the determination of the forces from the
particles.  However, we suggest that the recurrent cycle of modes also
develops here because of impedance changes caused by resonance
scattering in the inner disk.

\begin{figure}[t]
\begin{center}
\includegraphics[width=\hsize]{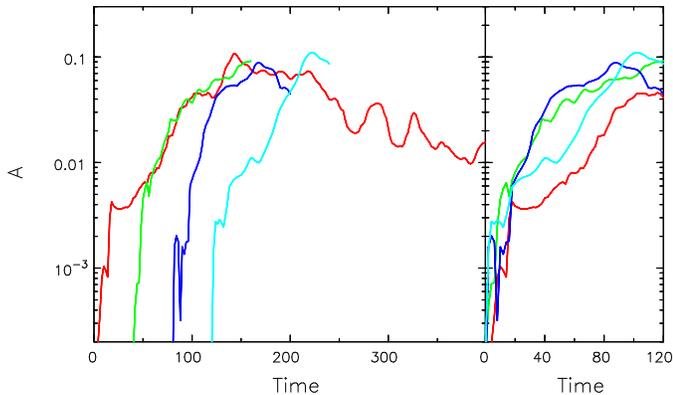}
\end{center}
\caption{Amplitude evolution of the $m=3$, $45^\circ$ trailing
  logarithmic spiral transform of the particle distribution in
  experiments with $N=2 \times 10^8$ particles.  The red curve shows
  the evolution in the model used in Fig.~\ref{fig.heat3D}, while the
  green, blue, and cyan curves show the behavior when the particles
  from that model have their azimuthal coordinates randomized after
  two, four, and six disk rotations, respectively.  The right panel
  shows the same lines shifted to a common start time.}
\label{fig.ampl20M}
\end{figure}

In our large-$N$ experiments, swing-amplification of the initial
low-amplitude shot noise created mild in-going spiral responses that
are fully absorbed at the ILR causing important, though small,
localized changes to the impedance of the disk.  Subsequent responses
to noise were partially reflected at these features, and the partial
feedback boosted later activity to a somewhat higher amplitude.  As
found by \citet{Sell12}, several cycles of scattering by
swing-amplified responses to shot noise were required before the
impedance changes wrought by the absorption of these transient wave
packets could give rise to reflections that were sufficiently strong
to trigger instabilities.  Hence, the increasing delay to the start of
visible spiral activity and disk heating reported in
Figs.~\ref{fig.heat2D} and \ref{fig.heat3D}.

Fig.~\ref{fig.ampl20M} presents further evidence in support of this
picture.  Each curve shows the amplitude of the $m=3$, $\cot\alpha=1$
component of the transform
\begin{equation}
A(m,\alpha,t) = {1 \over N} \sum_{j=1}^N \exp[im(\phi_j + \cot\alpha \ln R_j)],
\label{eq.logspi}
\end{equation}
where $(R_j,\phi_j)$ are the cylindrical polar coordinates of the
$j$-th particle at time $t$, and $\alpha$ is the pitch-angle of an
$m$-arm logarithmic spiral, with positive values for trailing spirals.
We have smoothed the curves in time using a running average of 10
consecutive values.

The red curve in Fig.~\ref{fig.ampl20M} shows the time evolution of
the amplitude, $|A|$, in our largest simulation ($N=2 \times 10^8$
particles).  The other lines show the amplitude evolution in three
further experiments, each derived from that used for the red line by
restarting from randomized coordinates of all the particles after two,
four and six rotations.  We merely rotated the radius vector to each
particle through a random angle, while preserving the radius, $R$, and
the polar velocity components $(v_R,v_\phi)$.  This procedure clearly
destroyed all non-axisymmetric structure at that moment, and reduced
the amplitude back to the shot noise level, while leaving the other
parts of the phase space structure intact.  It is clear that the green
and blue curves rise more quickly than does the red, indicating that
the models scrambled after two and four rotations possess more
vigorous instabilities than did the original.  

In the cases of the models restarted at $t=40$ and at $t=80$, shown by
the blue and green curves, we find the time evolution over the first
30 dynamical times is quite well fitted by two exponentially growing
modes, with growth rates in the range $0.07 \leq \gamma \leq 0.11$ and
both modes rotate slowly enough to possess ILRs, although the
precision of our frequency estimates is limited by the brevity of the
period of linear growth.  This contrasts with the original run, shown
by the red curve, where many frequencies are present in the evolution
to $t=40$ that appear to be mostly responses to swing-amplified shot
noise.

\Ignore{
Restarted at 40 (run4502)
 mode     Real part      Imaginary part
  1    1.1091+-0.1125 + (0.0856+-0.0224)i
  2    0.9030+-0.0485 + (0.1084+-0.0203)i

Restarted at 80 (run4502)
 mode     Real part      Imaginary part
  1    1.6465+-0.0171 + (0.0709+-0.0071)i
  2    0.8327+-0.0144 + (0.0953+-0.0168)i
}

Not only does this result show that the prior evolution has altered
the phase space structure in such a way as to provoke instability, but
the growing disturbances owe nothing to pre-existing density
variations in the disk, since they were all destroyed by randomizing
the azimuths of all the particles.  It is therefore inconsistent with
other possible explanations for the continued activity that invoke
non-linear effects \citep{DVH13} or mode coupling \citep{Tagg87,
  Fuch05}.

The cyan curve does not rise as rapidly as do the blue and green
curves because it was restarted at $t=120$ in the original model when
the disk had heated somewhat, especially in the inner parts.  The
vigor of instabilities in the inner disk at later times will be
reduced by the increased random motion, while those farther out in the
disk would grow more slowly anyway because the dynamical time scale is
longer.

\begin{figure}[t]
\begin{center}
\includegraphics[width=\hsize]{act.1.0.ps}
\smallskip
\includegraphics[width=\hsize]{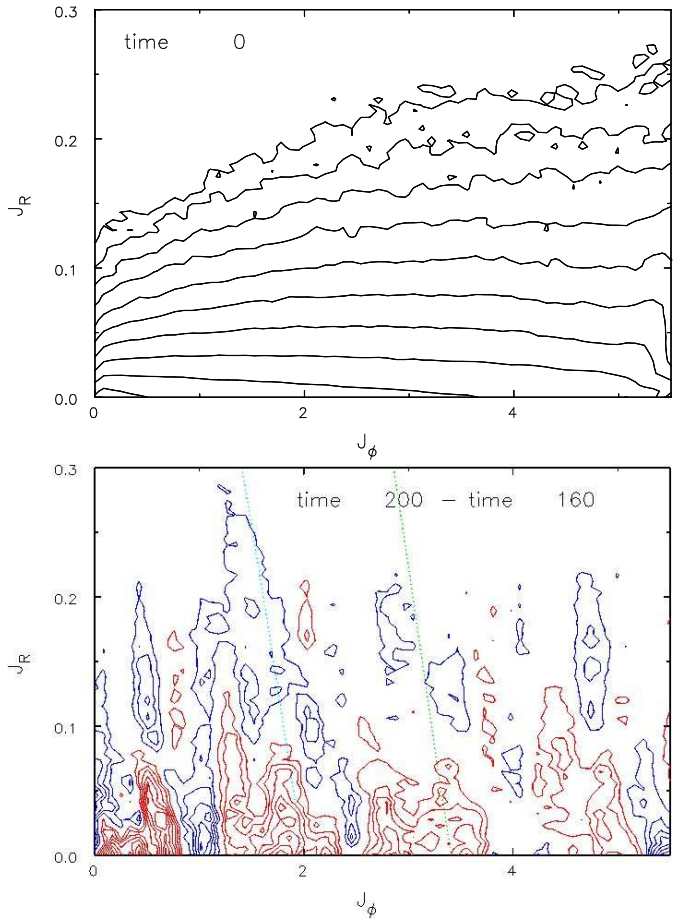}
\end{center}
\caption{Upper panel: the density of particles in the space of the
  actions $(J_\phi, J_R)$ at $t=0$ in the $N = 2 \times 10^6$ particle
  experiment.  The density maximum is near the origin, and the contour
  spacing is logarithmic; the third contour is at $0.1$ and the
  seventh is at $10^{-3}$ of the maximum.  Lower panel: changes to the
  density of particles on the same axes between times 160 and 200 in
  same simulation.  Contours in this panel are linearly spaced, and
  are colored blue where the density has increased, red where the
  density decreased.  The dotted lines indicate the slopes of ILRs in
  this plane for arbitrarily chosen frequencies; green is for $m=3$
  and cyan for $m=4$.}
\label{fig.actall}
\end{figure}

\subsection{Limiting Amplitude}
\label{sec.saturate}
A mode that grows exponentially at small amplitude must saturate as
the second- and higher-order terms, which are neglected in linear
analyses, become important.  The equilibrium state is unchanged to
first order, but it can be altered by the higher-order terms.

During the period when linear theory is adequate, the deflections of
the stellar motions caused by the growing potential perturbation must
reinforce the perturbed density, else the mode would not grow.
However, the orbital deflections of the particles change at finite
amplitude; in particular, horseshoe orbits appear near corotation.
\citet{SB02} argued that the maximum amplitude of a spiral is limited
by the widening horseshoe region where stars are driven away from,
instead of toward, the density maximum.  This change kicks in suddenly
because the exponentially growing disturbance density is linearly
dependent upon in the disturbance potential, but the width of the
horseshoe region grows as its square-root \citep{SB02}.  Empirically,
the relative overdensity in the disturbance reaches some 20\% or 30\%
before this behavior terminates growth of a linear mode.  The wave
then begins to decay about as rapidly as it grew \citep{SB02}, and all
the wave action, \ie\ angular momentum, stored in the disturbance
\citep{LBK72} is carried away from corotation at the group velocity
\citep{Toom69}.

\subsection{Recurrent Cycle}
Once the distribution function, hereafter DF, has become sufficiently
non-smooth to provoke one instability, further instabilities can give
rise to later spirals.  Each is a cavity mode with reflections off
corotation, where it is amplified, and some inner radius where
recent disturbances have caused abrupt changes to the impedance of
the disk.

Evidence for this picture is presented in Fig.~\ref{fig.actall}.  The
upper panel shows the density of particles in the space of the actions
at time 0, while the lower panel shows the change in the density of
particles between times 160 and 200 in the same space.  The azimuthal
action, $J_\phi \equiv L_z$ in an axisymmetric potential, while the
radial action, $J_R$, is a measure of the amount of in-and-out motion
of the particle.  At the initial moment, the density of particles in
this space is quite smooth, declining steeply with increasing random
motion, and slowly with increasing angular momentum.  The differences
at later times reveal rising features with steep negative slopes
marked by the blue contours that indicate movement of particles to
higher $J_R$ and slightly lower $J_\phi$, which is characteristic of
ILR scattering.  The dotted lines, which have similar slopes to each
other and to the ridges, indicate the slopes of two fiducial ILRs for
arbitrarily chosen pattern speeds.

The increases in $J_R$ over this short time period are localized at a
small number of ILRs.  The consequent abrupt changes to the impedance
of the disk cause traveling waves to be sufficiently strongly
reflected that the system can support fresh standing waves that are
amplified at corotation.  As reflections at impedance changes are
partial, some wave action continues to the ILR of the mode, where
further resonant scattering occurs.  Over time, the generally rising
level of random motion makes the disk less able to support and amplify
coherent waves and activity eventually dies away -- unless some
cooling is applied.

\begin{figure*}[t]
\begin{center}
\includegraphics[width=.6\hsize,angle=270]{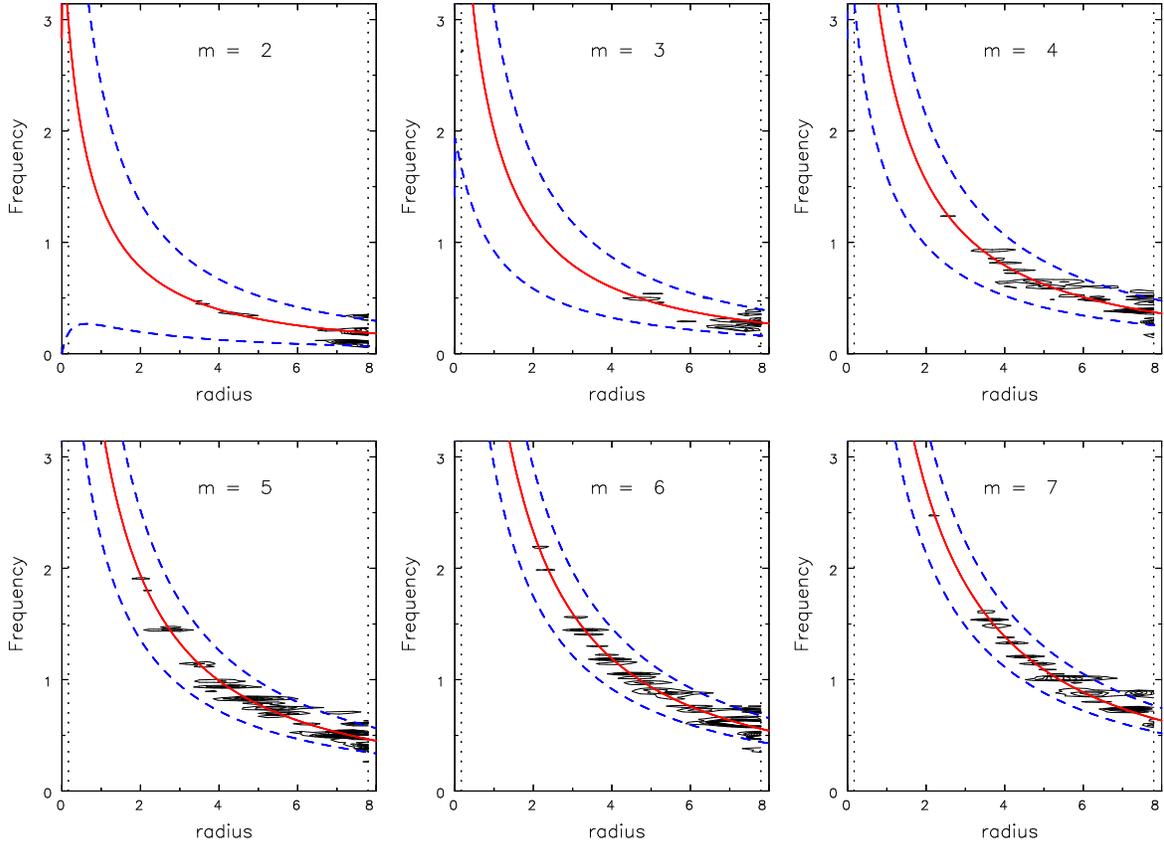}
\end{center}
\caption{Power spectra from the full duration of the low-mass disk
  model ($f=0.2$).  Notice that there is little power in this model at
  $2 \leq m \leq 3$ and most power is for $m \geq 4$, whereas sectoral
  harmonics $m>4$, not shown in Fig.~\ref{fig.pspectra}, had very
  little power in the disk with $f=0.4$.}
\label{fig.lspectra}
\end{figure*}

\section{Other recent work}
\label{sec.discuss}
\citet{Gran12a, Gran12b}, \citet{Baba13}, and \citet{Roca13} presented
a view of the behavior of spirals in their simulations that differs
from ours in at least two substantial respects.  We do not question
their simulation results, which are consistent with most other work,
but we take issue with their interpretation.

\subsection{Superposed Modes}
First, these authors argued that the spirals in their simulations of
generally rather low-mass disks appeared to shear continuously.  In
particular, they almost corotated with the stars at all radii, making
them qualitatively different from the usual density wave assumption
\citep[\eg][]{BT08}.

We have run simulations with $N = 2 \times 10^6$ particles and half
the disk mass of those presented above, \ie\ with $f=0.2$, and observe
multi-arm spiral features that appear to shear as these authors
describe.  However, power spectra, shown in Fig.~\ref{fig.lspectra}
again reveal coherent waves that qualitatively resemble those in the
more massive disk.

Most spiral activity in this low-mass disk occurs with $m \geq 4$ and
there is little power for sectoral harmonics $m \leq 3$.  Furthermore,
waves with higher angular periodicities have smaller radial extents
because the Lindblad resonances lie closer to corotation.  Higher
multiplicity spiral patterns are preferred in lower-mass disks for
reasons that are well understood (\SC).  The yardstick for dynamical
instabilities,
\begin{equation}
\lambda_{\rm crit} = {4\pi^2 G\Sigma \over \kappa^2}
\end{equation}
\citep{Toom64}, decreases with the disk mass (for a fixed rotation
curve).  Swing amplification is strongest where the wavelength of an
$m$-armed disturbance around its corotation circle
\begin{equation}
{2\pi R_{\rm  CR} \over m} \sim 2\lambda_{\rm crit} \quad\hbox{or}\quad
m \sim {R_{\rm CR} \kappa^2 \over 4\pi G\Sigma},
\end{equation}
\citep{GLB65, JT66, Toom81}.  Thus lower disk mass fractions, for
which $\lambda_{\rm crit}$ is shorter, favor higher multiplicity
patterns.

Fig.~\ref{fig.lspectra} reveals that apparently shearing patterns are
simply the result of the superposition of multiple separate modes.
The co-existence of two or more waves rotating at different angular
rates necessarily produces a shearing density ridge, provided those
waves with lower frequency have peak amplitudes at a larger
radii.\footnote{For an animation showing how this is possible, see
  {\tt http://www.physics.rutgers.edu/$\sim$sellwood/spirals.html}.}
Because these higher-multiplicity spiral modes have a smaller radial
extent than those in the high mass disk, the spectra must be computed
from long time periods of evolution with frequent analyses to reveal
the underlying modal behavior.

\subsection{Radial Mixing}
The authors of the papers cited at the start of this section offer a
description of the mechanism for radial mixing, which still occurred
in their simulations, that differs from the standard view \citep{SB02,
  Rosk12, Solw12} that {\it requires\/} stars to move through the
pattern, albeit slowly.

\begin{figure}[t]
\begin{center}
\includegraphics[width=.8\hsize]{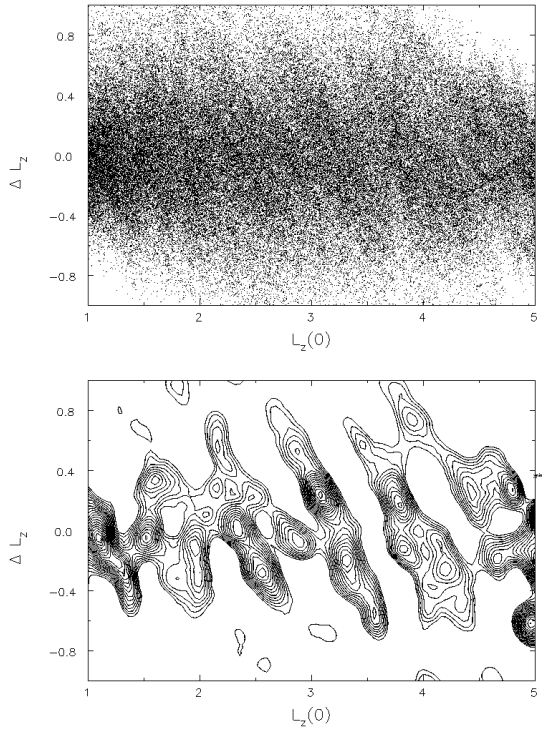}
\end{center}
\caption{Change in angular momentum, $\Delta L_z$, {\it vs.}  initial
  angular momentum, $L_z$ for particles in the low-mass disk
  simulation.  The upper panel shows a representative fraction of the
  particles while the lower panel shows contours of higher than
  average density in the same plane.}
\label{fig.migrate}
\end{figure}

In fact, the standard view still holds, as shown in
Fig.~\ref{fig.migrate}, which plots $\Delta L_z = L_z(400) - L_z(0)$
versus initial $L_z(0)$ for the particles in the low mass disk.  The
$\Delta L_z$ values display similar angled streakiness to that
reported by \citet{SB02} and by \citet{Solw12} that results from
distinct modes, even though the individual patterns may appear to be
shearing.

The angular momentum of a particle on a horseshoe orbit changes
substantially; those inside corotation gain enough angular momentum to
cross the resonance, while those outside lose a similar amount.  The
particles experience these changes once only for each transient spiral
mode because the disturbance has a large amplitude for less than half
a horseshoe orbit period \citep{SB02}.  Thus the dominant changes to
the angular momenta of particles near corotation of a single wave are
for low angular momentum particles to gain and high angular momentum
particles to lose, giving rise to a single angled streak in this plot.
The multiple streaks visible in Fig.~\ref{fig.migrate} arose from
multiple coherent waves that each scattered particles across its own
corotation resonance, which requires the wave to have had a coherent
frequency and to have saturated as described in \S\ref{sec.saturate}.

It should be noted that where linear perturbation theory holds,
\ie\ for small amplitude disturbances away from resonances, the
response of the disk particles when multiple waves are present can be
computed separately for each wave \citep[see \eg][pp 188-190]{BT08},
with the net response being the sum of the separate linear responses.
Thus resonant scattering by one wave is unaffected by the co-existence
of other waves, unless resonances were to overlap.

\begin{figure}[t]
\begin{center}
\includegraphics[width=.8\hsize]{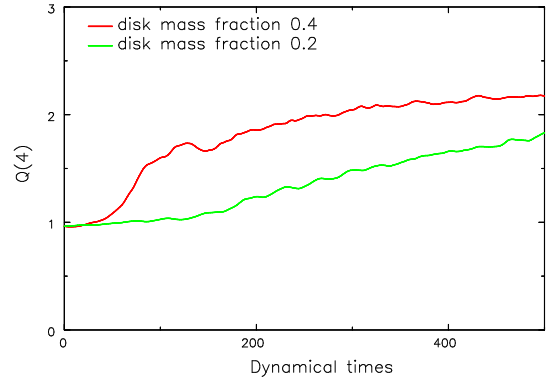}
\end{center}
\caption{Time evolution of $Q$ at $R=4$ in two 3D models with the same
  $N=2 \times 10^6$, but different active mass fractions.  The lighter
  disk (green) has $f=0.2$, which is half the mass fraction of those
  shown in Fig.~\ref{fig.heat3D}, and reproduced in red.}
\label{fig.Qmfrac}
\end{figure}

\subsection{Slower Disk Heating}
\label{sec.slowheat}
\citet{Fuji11}, and the authors of the papers cited at the beginning
of this section, also reported that spirals heat the disk more slowly
than the rate reported in \SC.  Fig.~\ref{fig.Qmfrac} shows that the
heating rate is indeed lower in disks with smaller active mass
fractions.  The reason for the difference is clear: lower-mass disks
support patterns of higher multiplicity (Fig.~\ref{fig.lspectra}),
which therefore have Lindblad resonances closer to corotation.  Thus
the angular momentum transferred outward over the shorter radial
extent of these multi-arm patterns releases less energy into
non-circular motion (see Fig.~\ref{fig.lindblad}).

\subsection{Heavy Particles}
\citet{DVH13} employed $N=10^8$ particles in their simulations of a
low-mass disk, and experimented with the consequences of adding a few
co-orbiting, heavy particles.  They ran a model with no heavy
particles for about 2.5 disk rotations during which no visible change
occurred, but it is likely that non-axisymmetric disturbances were
growing that would have appeared had they continued the evolution.
(We note that our experiment with a low-mass disk employing merely $2
\times 10^6$ particles manifested no visible activity, and little
heating for perhaps eight rotations, as shown by the green curve in
Fig.~\ref{fig.Qmfrac}.)

\citet{DVH13} showed that starting similar models with a sprinkling of
heavy particles provoked almost immediate multi-armed spiral activity.
In one case, they introduced a single perturbing particle that
produced a one-armed spiral response and then removed it again one
full disk rotation after the start.  Four disk rotations after the
perturber was removed, they showed that the disk continued to manifest
multiple spiral arms, which they attributed to the non-linear
evolution of the original disturbance.  We suggest instead that their
system developed self-excited instabilities due to changes to the
originally smooth DF of the disk.  Instabilities, which grew out of
the initial noise in our experiments were, in their experiment, seeded
by the response to the massive co-orbiting particle they introduced,
enabling visible self-excited activity immediately thereafter.

Of course, the heavy particles employed by \citet{DVH13} were intended
to mimic the molecular cloud complexes possessed by real galaxy disks.
\citet{JT66} showed that a massive, co-orbiting perturber provoked a
spiral response in the surrounding stellar disk having an amplitude
that scaled with the mass of the perturber.  We have preferred to
avoid further complicating the dynamical picture with this added
realism, as we wish to understand how purely collisionless disks can
develop and support recurring spiral patterns.  However, if spirals
can develop as self-excited instabilities, as we have argued here,
then the role of heavy clumps in the disk is probably not fundamental
to the origin of spiral patterns \citep[\cf][]{Toom90}.

\section{Conclusions}
We have presented evidence that spiral activity in simulations of
cool, unbarred, collisionless stellar disks results from a recurrent
cycle of transient spiral modes of spiral form
(Fig.~\ref{fig.spirals}).  We describe them as modes because they
start as linear instabilities that grow exponentially even from very
low-amplitude (Fig.~\ref{fig.ampl20M}) before saturating and decaying.

Growing modes are standing wave oscillations that have positive feedback
to cause instability.  We argue that the growing wave train reflects
off corotation, where it is swing-amplified, and again at some inner
radius, where the distribution function has been modified by previous
disturbances in the disk.  Scattering of particles at resonances
causes localized heating over a moderately narrow range of angular
momenta (Fig.~\ref{fig.actall}), which introduces abrupt changes to
the impedance experienced by traveling waves.  Such changes cause
partial reflection of a subsequent wave, allowing a standing wave, or
unstable mode to develop.

Spiral instabilities saturate as a result of the onset of horseshoe
orbits that appear as the relative overdensity near corotation
approaches $\sim 20$\%, as originally proposed by \citet{SB02}.  Once
growth is halted by the dispersal of the overdensity at corotation,
the wave action stored in the remaining disturbance propagates away
from corotation until it is absorbed by wave-particle interactions
that cause further localized heating.  Thus, the instability cycle is
able to repeat.

The repeated scattering of particles at different locations leads to a
general rise of random motion over the disk that weakens its ability
to support further coherent waves, and activity gradually fades on a
time-scale of some twenty disk rotations.  We have also shown that
this timescale is longer in low-mass disks because the multi-arm
patterns that are dynamically favored in this case transport angular
momentum over a shorter radial distance and, therefore, release less
energy into random motions (\S\ref{sec.slowheat}).

While we recognize that we have not substantiated all the details, we
have presented considerable evidence to support our broad picture.  In
particular, we find the apparent rapidly changing spirals result from
the superposition of a small number of relatively long-lived coherent
waves (Fig.~\ref{fig.pspectra}); the phase coherence and large
limiting amplitude of these waves are most naturally accounted for by
unstable modes.  The evolution of each disturbance creates the seeds
for a fresh instability, since we find more vigorous growth in
simulations that are restarted after scrambling only the azimuthal
phases of the particles (Fig.~\ref{fig.ampl20M}).  Not only does this
result support our picture, but it shows that the activity in the
simulations owes nothing to pre-existing density structures, that were
erased by scrambling.  Fig.~\ref{fig.actall} presents evidence of
resonance scattering, which we argue changes the impedance of the disk
to traveling waves, thereby creating features that cause partial
reflection of the waves, allowing fresh cavity modes to develop.

Much of our picture builds on previous work by many authors: the
dispersion relation for spiral waves \citep{LS66}, their group
velocity \citep{Toom69}, swing-amplification \citep{GLB65, JT66,
  Toom81}, resonance scattering \citep{LBK72, Mark74}, feedback loops
\citep{Mark77, Toom81}, global mode analyses \citep{Zang76, Kaln77,
  ER98}, and horseshoe orbits at corotation \citep{SB02}.  We could
not have reached our present level of understanding without all of
these contributions, yet our picture is distinct from any previous
suggestion.  In particular, we argue that the assumption of a smooth
distribution function, which many authors regard as the natural
starting point, fundamentally discards the spiral baby with the
bathwater!

Direct observational tests of the generating mechanism for spirals are
difficult to devise.  Evidence for density waves was summarized in the
introduction, but does not help to distinguish between rival theories
for their origin.  However, analysis of the complete phase-space
information for a sample of solar neighborhood stars \citep{Sell10}
revealed evidence for a resonant scattering feature, of the kind
illustrated in Fig.~\ref{fig.actall}, which supports the picture we
present here.  Furthermore, the existence of such a feature is
inconsistent with other leading theories for the origin of spiral
patterns \citep[\eg][]{Bert89, Toom90}.

\section*{Acknowledgments}
We thank the anonymous referee for a helpful report.  This work was
supported by NSF grant AST-1108977 to J. A. S., and hospitality from
APCTP during the 7th Korean Astrophysics Workshop is gratefully
acknowledged.  R. G. C. thanks the Canadian NSERC and the Canadian
Institute for Advanced Research for support.

\end{document}